\documentclass[num-refs]{wiley-article}

\usepackage{siunitx}
\usepackage[T1]{fontenc}

\usepackage[version=3]{mhchem} 

\usepackage{balance}
\usepackage{amsmath}

\papertype{Original Article}
\paperfield{Journal Section}

\title{Diffuson-Dominated Thermal Transport Crossover from Ordered to Liquid-like Cu$_3$BiS$_3$:The Negligible Role of Ion Hopping}

\author[1\authfn{3}]{Jincheng Yue}
\author[2,3,4\authfn{3}\authfn{1}]{Jiongzhi Zheng}
\author[5\authfn{3}]{Xingchen Shen}
\author[5]{Krishnendu Maji}
\author[6]{Chun-Chuen Yang}
\author[7,8]{Shuyao Lin}
\author[9]{Pierric Lemoine}
\author[5]{Emmanuel Guilmeau}
\author[1\authfn{1}]{Yanhui Liu}
\author[1,10\authfn{1}]{Tian Cui}

\affil[1]{Institute of High Pressure Physics, School of Physical Science and Technology, Ningbo University, Ningbo, 315211, China}
\affil[2]{Energy Technologies Area, Lawrence Berkeley National Laboratory, Berkeley, CA 94720, USA}
\affil[3]{Thayer School of Engineering, Dartmouth College, Hanover, New Hampshire, 03755, USA}
\affil[4]{Department of Mechanical and Aerospace Engineering, The Hong Kong University of Science and Technology, Clear Water Bay, Kowloon, 999077, Hong Kong}
\affil[5]{CRISMAT, CNRS, Normandie Univ, ENSICAEN, UNICAEN, 14000 Caen, France}
\affil[6]{Department of Physics, National Central University, Chung-Li District, Taoyuan City, 320317, Taiwan}
\affil[7]{Technische Universit\"{a}t Wien, Institute of Materials Science and Technology, Vienna, A-1060, Austria}
\affil[8]{Link\"{o}ping University, Department of Physics, Chemistry, and Biology (IFM), Link\"{o}ping, SE-58183, Sweden}
\affil[9]{Université de Lorraine, CNRS, IJL, F-54000 Nancy, France}
\affil[10]{State Key Laboratory of Superhard Materials, College of Physics, Jilin University, Changchun 130012, China}

\corraddress{
    Institute of High Pressure Physics, School of Physical Science and Technology, Ningbo University, Ningbo 315211, China 
}
\corremail{jzheng4@lbl.gov; liuyanhui@nbu.edu.cn; cuitian@nbu.edu.cn}

\begin{document}

\begin{frontmatter}
\maketitle

\begin{abstract}
Fundamentally understanding lattice dynamics and thermal transport behavior in liquid-like, partially occupied compounds remains a long-standing challenge in condensed matter physics. Here, we investigate the microscopic mechanisms underlying the ultralow thermal conductivity in ordered/liquid-like Cu$_3$BiS$_3$ by combining experimental methods with first-principles calculations. We first experimentally synthesize and characterize the ordered structure and liquid-like, partially Cu-atom occupied Cu$_3$BiS$_3$ structure with increasing temperature. We then combine self-consistent phonon calculations, including bubble-diagram corrections, with the Wigner transport equation, considering both phonon propagation and diffuson contributions, to evaluate the anharmonic lattice dynamics and thermal conductivity in phase-change Cu$_3$BiS$_3$. Our theoretical model predicts an ultralow thermal conductivity of 0.34 W·m$^{-1}$·K$^{-1}$ at 400 K, dominated by diffuson contributions, which accurately reproduces and explains the experimental data. Importantly, the machine-learning-based molecular dynamics (MD) simulations not only reproduced the partially Cu-atom occupied Cu$_3$BiS$_3$ structure with the space group $Pnma$ but also successfully replicated the thermal conductivity obtained from experiments and Wigner transport calculations. This observation highlights the negligible impact of ionic mobility arising from partially occupied Cu sites on the thermal conductivity in diffuson-dominated thermal transport compounds. Our work sheds light on the minimal impact of ionic mobility on ultralow thermal conductivity in phase-change materials. It demonstrates that the Wigner transport equation accurately describes thermal transport behavior in partially occupied phases with diffuson-dominant thermal transport.

\end{abstract}
\end{frontmatter}

\section{Introduction}
\par Ultra-low lattice thermal conductivity ($\kappa_L$) in solid-state materials is crucial for advancing energy conversion technologies~\cite{wachsman2011lowering, dresselhaus2007new}, improving fuel cell performance~\cite{niedziela2019selective}, and optimizing various thermoelectric applications in the fields of thermal barrier coatings~\cite{padture2002thermal}, waste heat recovery~\cite{roychowdhury2018soft, li2023overdamped}. However, understanding and measuring lattice thermal conductivity is particularly challenging for materials undergoing phase transitions~\cite{chen2019thermal}. These materials are widely available and have been extensively used in various fields, including thermoelectrics~\cite{liu2013ultrahigh}, solid-state memory~\cite{wuttig2007phase}, and switches~\cite{el2014low}. Therefore, a fundamental understanding of the origins of ultra-low thermal conductivity through a comprehensive investigation of the relationships between transport phenomena, lattice dynamics, and structural configurations is of great significance. This understanding unveils critical insights and establishes foundational principles for the innovation and discovery of advanced material systems~\cite{hanus2021uncovering}.

\par Within the family of phase transition materials, Cu-based chalcogenides stand out due to their complex phase transition behaviors and outstanding thermoelectric properties~\cite{liang2019phase, ghata2024exploring}. Cu-based chalcogenides often undergo complex lattice distortions and significant changes in thermal transport properties during phase transitions, making them a key platform for exploring new thermal conduction mechanisms. For instance, Liu et al.~\cite{liu2012copper} reported a significant reduction in the thermal conductivity of Cu$_2$Se, attributed to its second-order phase transition, which led to a notable enhancement of the thermoelectric figure of merit ($zT$) from 0.6 to 2.3. The recent synthesis of copper-rich wittichenite Cu$_3$BiS$_3$~\cite{wei2018wittichenite, jia2023cu3bis3, maji2024three}, which exhibits intrinsically low lattice thermal conductivity, is particularly noteworthy. This material provides a compelling platform to investigate structural features—such as atomic coordination and cationic networks—that influence thermal transport. Meanwhile, Maji et al. demonstrate that the 3-fold coordination of copper, combined with its strong anisotropic vibrations, serves as the driving force behind the ultra-low thermal conductivity in Cu$_3$BiS$_3$~\cite{maji2024three}. Despite extensive researches on Cu$_3$BiS$_3$, its phase transition behavior remains unclear, and the mechanisms underlying its ultra-low thermal conductivity across different phases have yet to be fully elucidated.

\par Historically, thermal transport in crystalline materials has been well described by the phonon-gas model, which treats phonons as quasiparticles propagating over long distances within the framework of the conventional Boltzmann transport equation (BTE)~\cite{toberer2011phonon, xia2020particlelike}. Within the conventional BTE framework, point defects~\cite{qu2011thermal}, anharmonicity~\cite{yue2024role, yue2023pressure}, and bond strength variations~\cite{ma2021mixed, yue2024ultralow} act as phonon scattering sources, reducing mean free paths and thereby lowering lattice thermal conductivity. However, certain complex crystalline materials exhibit ultra-low thermal conductivities that cannot be fully explained by the phonon-gas model~\cite{mukhopadhyay2018two}. The conventional BTE assumes well-separated phonon branches, and it therefore fails when this separation collapses, or when off-diagonal velocity terms become non-negligible—situations common in strongly anharmonic or disordered materials~\cite{allen1989thermal, zheng2022anharmonicity}. To address the limitations of the conventional BTE, Simoncelli et al.~\cite{simoncelli2019unified} developed a unified theory of thermal transport by reformulating the Boltzmann Transport Equation within the Wigner formalism. Specifically, the Wigner transport equation~\cite{simoncelli2022wigner} describes thermal transport via the phonon velocity operator, wherein the diagonal and off-diagonal elements (diffuson described by Allen and Feldman~\cite{allen1989thermal, allen1999diffusons}) capture both the particle-like propagation of phonons and the wave-like tunneling between eigenstates. The wave-like tunneling arises from the overlap of phonon modes with similar frequencies—substantially broadened by anharmonicity—thereby facilitating thermal energy transfer between modes. The Wigner transport equation has been successfully applied to address the underestimation of thermal conductivity within the BTE framework for strongly anharmonic systems, such as Cu$_{12}$Sb$_4$S$_{13}$\cite{xia2020microscopic}, La$_2$Zr$_2$O$_7$\cite{simoncelli2019unified}, Cs$_2$AgBiBr$_6$~\cite{zheng2024unravelling}.

\par Another challenge in evaluating the thermal conductivity of phase-change materials is the potential occurrence of a partial-crystalline partial-liquid state (PCPLS), driven by the liquid-like migration of Cu ions, as observed in superionic systems such as Ag$_9$GaSe$_6$\cite{lin2017high}, AgCrSe$_2$\cite{wang2023anisotropic}, and Li$_2$S~\cite{zhou2018thermal}. It is also reported that the most representative $\alpha$-Cu$_2$S, with space group $Fm\bar{3}m$ and Pearson symbol cF208, exhibits a remarkably high degree of disorder, with 8 Cu ions distributed over 204 possible sites within the unit cell~\cite{cava1981mobile, ma2019cscu5s3}. 
The strikingly structural disorder suggests an exceptionally low activation barrier for Cu-ion migration, giving rise to liquid-like cationic dynamics. This remarkable ion mobility facilitates strong phonon scattering mechanisms, thereby reducing the specific heat via the suppression of transverse phonon modes~\cite{he2014high}. In contrast, Bernges et al.~\cite{bernges2022considering} conducted the first direct comparison of thermal and ionic transport in nine Ag$^+$ argyrodite compositions. Their results revealed that, despite the ionic conductivity varying by several orders of magnitude, the thermal conductivity remains largely constant, indicating that high ionic mobility is not essential for diffuson-mediated thermal transport. Similarly, Ghata et al.~\cite{ghata2024exploring} observed that with increasing temperature, the ionic conductivity of Cu$_7$PSe$_6$ rises significantly, while the thermal conductivity remains nearly constant, suggesting a negligible influence of ionic transport on thermal conductivity. Therefore, a fundamental understanding of the thermal transport mechanism in phase-transition and partially occupied materials remains an open question and an urgent need.

\par  In this work, we systematically investigate the anharmonic lattice dynamics and thermal transport behavior in ordered and elemental-partially occupied, liquid-like Cu$_3$BiS$_3$ by combining experimental and theoretical approaches. Experimentally, we synthesize and characterize the ordered  and liquid-like phases of Cu$_3$BiS$_3$, which adopt the space groups $P{2_12_12_1}$ and $Pnma$, respectively. We then combine first-principles calculations with machine-learning-based molecular dynamics simulations to accurately reproduce the partially Cu-atom occupied phase of Cu$_3$BiS$_3$ ($Pnma$) at high temperatures. Subsequently, we applied the state-of-the-art Wigner transport equation, accounting for both propagative and diffuson contributions, to calculate the thermal conductivity of the $Pnma$ and $P{2_12_12_1}$ phases of Cu$_3$BiS$_3$, achieving excellent agreement with experimental results. To investigate the impact of ion hopping on thermal transport at high temperatures in Cu$_3$BiS$_3$, we calculate the lattice thermal conductivity ($\kappa_L$) using the machine-learning-based Green–Kubo equilibrium molecular dynamics (GK-EMD) method~\cite{kubo1957statistical}, which explicitly accounts for all contributions to the heat-flux operator~\cite{marcolongo2016microscopic, carbogno2017ab}. We find that the predicted thermal conductivity accurately reproduces both the experimental data and the results obtained from the Wigner transport equation, indicating a negligible effect of ion mobility on thermal transport. Finally, we explore the microscopic origins of the ultralow thermal conductivity observed in both phases of Cu$_3$BiS$_3$. Our work not only highlights the negligible role of ionic mobility in thermal transport but also provides deeper insights into the microscopic mechanisms underlying the ultralow thermal conductivity observed in ordered and liquid-like Cu$_3$BiS$_3$ and related compounds. 

\section{Experimental crystal structures}

\begin{figure}[t]
\centering
\includegraphics[width=13cm]{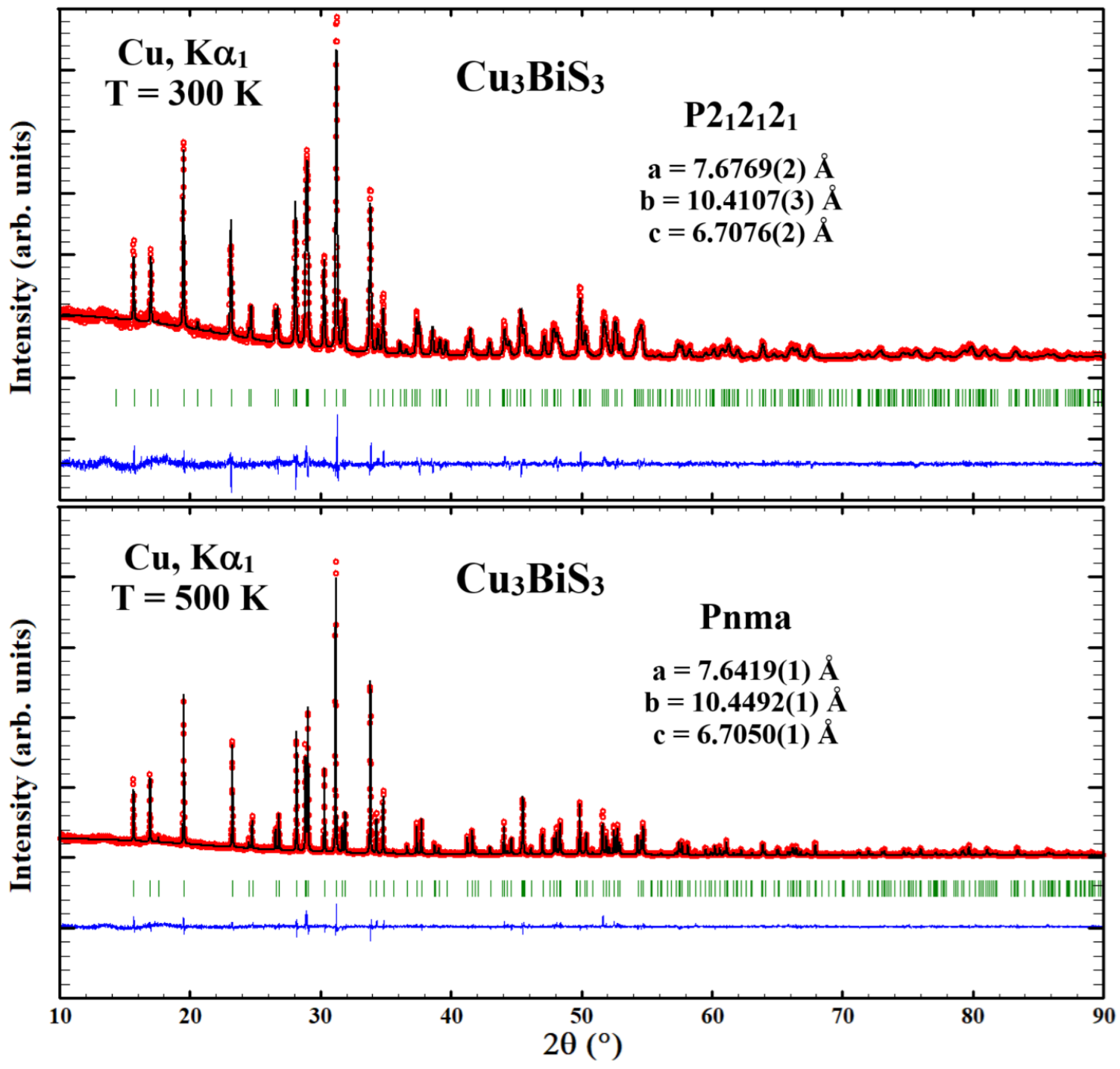}
\caption{ Rietveld refinement of the powder X-ray diffraction (PXRD) data recorded at 300 K and 500 K for the Cu$_3$BiS$_3$ sample and representations of the different crystal structures of Cu$_3$BiS$_3$ (space group $P{2_12_12_1}$ and $Pnma$. Purple, blue, and yellow colors represent Bi, Cu, and S, respectively.}
\end{figure}

\begin{figure}[t]
\centering
\includegraphics[width=14cm]{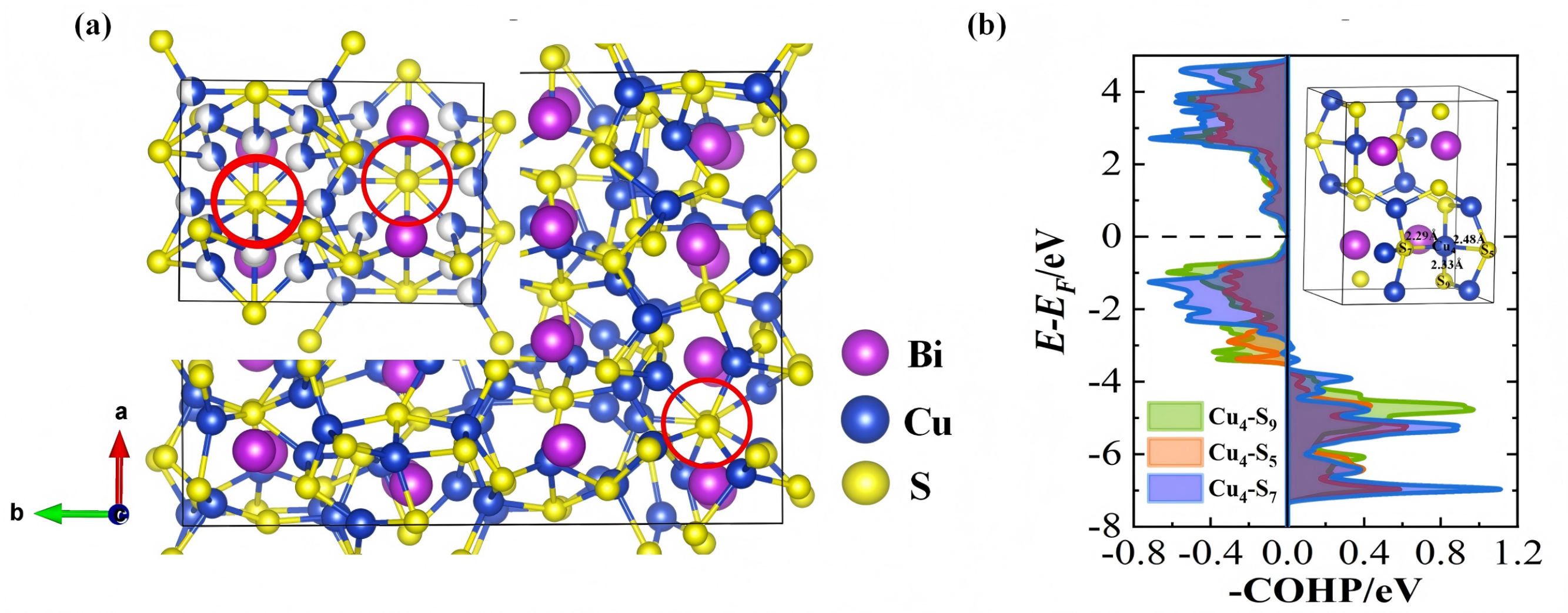}
\caption{ (a) Trajectory sampling plot of ab initio molecular dynamics simulation (AIMD) at 500 K. Inset: Crystal structure characterization with eight-fold coordination of S atom. (b) The comparison of summed crystal orbital Hamilton populations for different Cu–S bonds.}
\end{figure}

\par Rietveld refinement carried out using the Fullprof software package~\cite{rodriguez1993recent, roisnel2001winplotr}, leads to calculated patterns in excellent agreement with the powder X-ray diffraction (PXRD) data, as illustrated in Figure 1.
As evidenced by the well-defined refinement at 300 K (R$_p$ = 7.96, R$_{wp}$ = 10.3, R$_{exp}$ = 9.13 and $\chi^2$ = 1.28), the PXRD pattern is accurately indexed to the non-centrosymmetric space group $P{2_12_12_1}$, consistent with previously reported results~\cite{maji2024three, kocman1973crystal, makovicky1983phase}. Specifically, the crystal structure is characterized by Cu, Bi and S atoms located on three, one and three differents crystallographic sites, respectively, of general position, i.e. Wyckoff 4a. Each Bi atom is bonded to three S atoms at distances ranging between 2.565 Å and 2.614 Å (average Bi-S distance of 2.594 Å), forming a [BiS$_3$]$^{3-}$ trigonal pyramid.
Note that the lone pair (LP) of electrons of Bi$^{3+}$ cation very probably constitutes the missing apex of a tetrahedron [BiS$_3$LP]$^{3-}$. In contrast, each Cu atom is coordinated by three S atoms at distances ranging between 2.178 Å and 2.381 Å (average Cu-S distance of 2.285 Å), forming a distorted trigonal planar arrangement. The unit cell parameters determined from Rietveld refinement of the phase-pure sample, namely, a = 7.6769(2) Å, b = 10.4107(3) Å, and c = 6.7076(2) Å, are in agreement with density functional theory (DFT) calculations performed using the PBEsol functional~\cite{perdew2008restoring} (a = 7.4618(8) Å, b = 10.4705(9) Å, c = 6.3209(5) Å) as well as with previously reported results~\cite{jia2023cu3bis3, maji2024three}.
 
\par Upon further increasing the temperature, we observe a notable change in the XRD pattern (see Figure S1 in the Supporting Information) confirming previous results reported by Makovicky highlighting that Cu$_3$BiS$_3$ exhibits a structural transition from a RT orthorhombic ($P{2_12_12_1}$) structure to an intermediate modulated structure at 391.5 K and, above 463 K, a transition to a high temperature (HT) polymorph ($Pnma$)~\cite{makovicky1983phase}. Based on the structural refinments of the high-T PXRD patterns, we confirm an experimental phase transition from the orthorhombic space group $P{2_12_12_1}$ to the orthorhombic space group $Pnma$, as illustrated in Figure 1, even if the space group $Pna2_1$ is not excluded~\cite{makovicky1983phase}. This space group change from $P{2_12_12_1}$ at 300 K to $Pnma$ at 500 K is related to the extinction of the (hk0) reflections for which h = 2n+1, as exemplified with the (120) and (310) reflections observed at 20.6° and 36.1° in 2$\theta$ on the PXRD pattern recorded at 300 K and absent at 500 K. In addition, a broad peak observed near 2$\theta$ $\approx$ 13.2° does not match the known reflections of the Cu$_3$BiS$_3$ phase and is significantly wider than the main phase peaks. This suggests that it may arise from the sample environment or from a poorly crystallized impurity phase that could not be reliably identified. On the other hand, we have observed that CuBiS$_2$ exists in the $Pnma$ crystal system based on the structural data reported by Ref.~\cite{portheine1975refinement}. To rule out the possibility of CuBiS$_2$ as a minority or mis-indexed phase, we simulated powder-diffraction patterns for the three most plausible structures at the Cu, K$\mathrm{\alpha}$ wavelength: (i) Cu$_3$BiS$_3$ in $P{2_12_12_1}$ and $Pnma$, the two space groups obtained from our Rietveld refinements, and (ii) CuBiS$_2$ in $Pnma$, using the atomic coordinates reported by Portheine and Nowacki~\cite{portheine1975refinement} (see Figure S2 in the Supporting Information). The calculated pattern for CuBiS$_2$ displays an intense reflection at 2$\theta$ $\approx$ 12°, a feature that is completely absent in our experimental data, where the first detectable peak appears above 15°. Additional reflections predicted for CuBiS$_2$ (e.g., 2$\theta$ $\approx$ 18.3°, 26.9°, 31.7°) are also missing or mismatched in both position and intensity. These systematic discrepancies unambiguously exclude CuBiS$_2$ as a constituent phase in the investigated sample. To quantify the competition between ordered and liquid-like states from first principles, we have constructed a Cu–Bi–S phase diagram based on the convex-hull of DFT formation enthalpies (see Figure S3(a) in the Supporting Information). The diagram encompasses all computed compositions and explicitly resolves the two Cu$_3$BiS$_3$ polymorphs: the orthorhombic $Pnma$ phase and the $P{2_12_12_1}$ ordered variant. Notably, the $P{2_12_12_1}$ structure lies 3.16 meV/atom below $Pnma$, i.e. $\Delta E=E_{P2_12_12_1}-E_{Pnma}$ = –3.16 meV/atom, confirming its thermodynamic preference.

\begin{figure}[t]
\centering
\includegraphics[width=14cm]{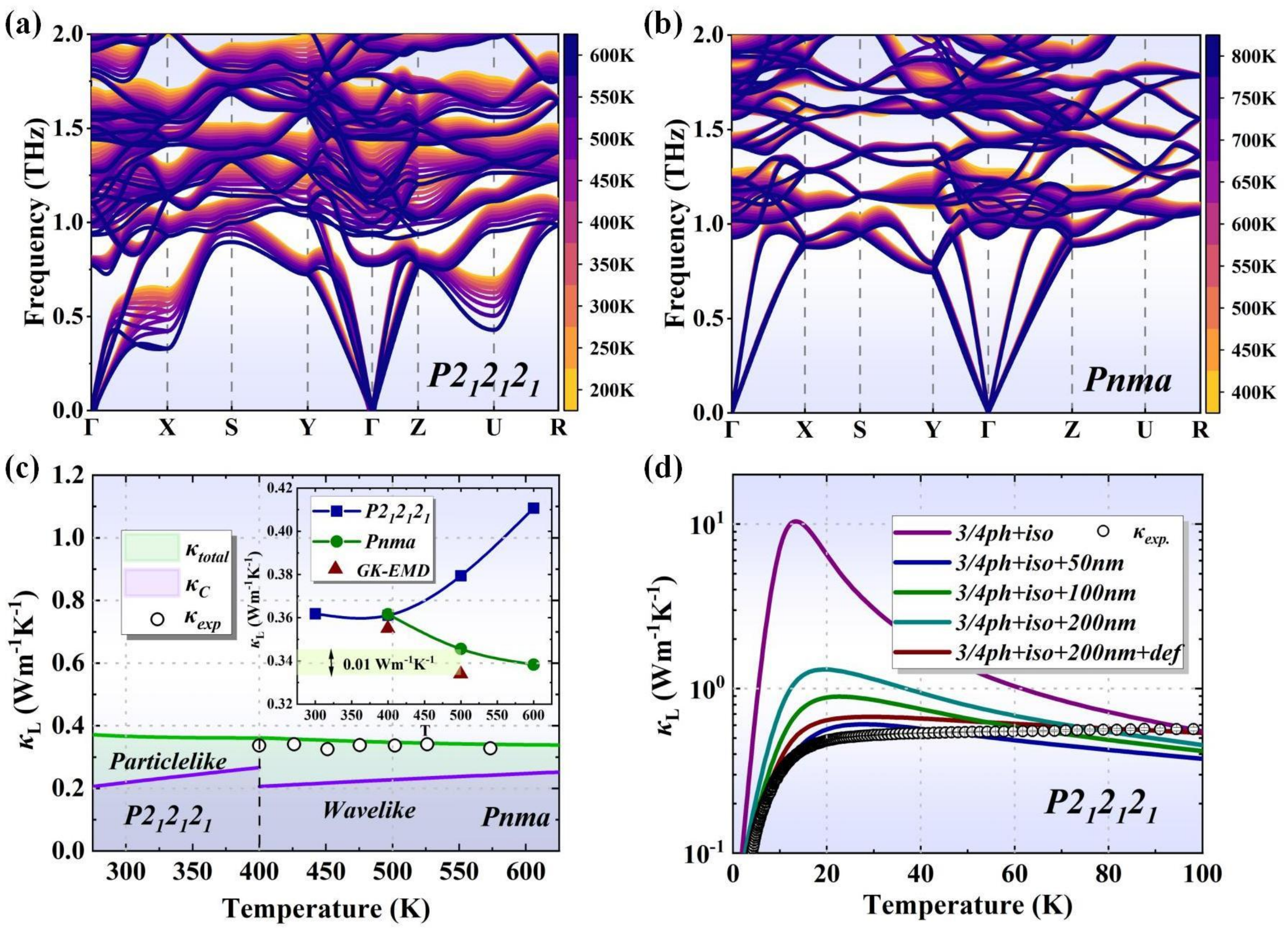}
\caption{ Finite-temperature phonon dispersions calculated using the SCPH method with bubble digram correction (SCPB method) for (a) $P{2_12_12_1}$ phase and (b) $Pnma$ phase. (c) Calculated temperature dependent averaged lattice thermal conductivity, including contributions from population and coherence conductivities, considering both 3ph and 4ph scattering processes. Illustration: Comparison of the thermal conductivity of the $P{2_12_12_1}$ phase (calculated from 300 to 600 K) and the $Pnma$ phase (calculated from 400 to 600 K) with the $Pnma$ phase obtained using the Green-Kubo method at 400 and 500 K. (d) Comparison of the total thermal conductivity, taking into account various scattering mechanisms such as three-, four-phonon, isotope, boundary, and point defect scattering, with experimental measurements from 2 to 100 K.}
\end{figure}

\par Interestingly, the distribution of Cu$^+$ ions within the framework of [BiS$_3$]$^{3-}$ units has been elucidated in the $Pnma$ phase (see Figure S3(b) in the Supporting Information), matching that observed in $\alpha$-Cu$_3$SbS$_3$~\cite{pfitzner1998disorder}. Specifically, single-crystal data recorded at 493 K reveal that Cu$^+$ ions exhibit disorder (see Figure S3(c) in the Supporting Information), being distributed over five distinct sites each being partially occupied, resulting in either trigonal-planar or tetrahedral coordination. To investigate this phenomenon, we conducted 20 ps ab initio molecular dynamics (AIMD) simulations at 400 K and 500 K (see Figure 2(a) and Figure S3(d-e) in the Supporting Information). Notably, the simulations successfully captured the diverse sites occupied by Cu$^+$ ions, confirming their mobility and aligning closely with our experimental observations. To elucidate the chemical environment of Cu$^+$ ions, we performed electronic structure calculations on the $Pnma$ phase, examining both bonding and antibonding states to gain deeper insights into its bonding characteristics. In the $Pnma$ phase, various Cu–S bonds exhibit distinct bonding characteristics (see Figure 2(b)); nevertheless, these bonds generally share common features. All Cu–S bonds exhibit occupied antibonding states below the Fermi level, contributing approximately 45\% to the overall antibonding character. Previous studies have attributed this behavior to the proximity of the Cu d-orbitals to the energy levels of the S p-orbitals, leading to strong p–d hybridization~\cite{gholami2024unlocking}. Such robust hybridization pushes the occupied antibonding states closer to the Fermi level, weakening the chemical bonds and promoting metavalent-like behavior~\cite{yue2024ultralow, yue2024hierarchical}. The combination of low integrated electron population and substantial antibonding contributions highlights the intrinsically weak bonding environment of Cu$^+$ within the structure, thus facilitating rapid Cu$^+$ ion transport~\cite{hong2019phase}. Furthermore, the significant presence of occupied antibonding states below the Fermi level notably enhances lattice anharmonicity, substantially influencing thermal transport, as discussed in subsequent sections. Also, such a weak bonding leads to larger mean square displacement (MSD) of Cu atoms compared to the other atoms (see Figure S4 in the Supporting Information). This is consistent with the experimental observation of the atomic displacement parameters of Cu atoms reported in Cu$_3$BiS$_3$~\cite{maji2024three}.

\section{Diffuson-dominated thermal transport}

\par To accurately capture the finite-temperature lattice dynamics, the anharmonic phonon frequencies of crystalline Cu$_3$BiS$_3$ are calculated using self-consistent phonon theory with bubble correction (SCPB)~\cite{zheng2024unravelling, tadano2022first} (see Figures 3(a) and 3(b)). Interestingly, from Figures 3(a) and 3(b), we find that in both the $P{2_12_12_1}$ and $Pnma$ phases, low-frequency phonons exhibit softening with increasing temperature, in significant contrast to previously reported results for other crystalline compounds, such as BaZrO$_3$~\cite{zheng2022anharmonicity}, TlInTe$_2$~\cite{pal2021microscopic}, and YbFe$_4$Sb$_{12}$~\cite{zheng2022effects}. Note that in strongly anharmonic materials, such as CsPbBr$_3$~\cite{tadano2022first}, KNbO$_3$ and NaNbO$_3$~\cite{kim2023exploring}, the negative energy shift from the bubble diagram~\cite{tadano2022first} is significant and imposes a non-negligible impact on both the theoretical phase transition temperature and the phonon linewidth. Specifically, considering only the energy shift from quartic anharmonicity leads to low-frequency phonon hardening in both the $P{2_12_12_1}$ and $Pnma$ phases of Cu$_3$BiS$_3$ (see Figure S5 in the Supporting Information). Therefore, incorporating the full explicit anharmonic vibrational effects, including both cubic and quartic anharmonicity, enables a more accurate description of phonon softening, which significantly influences the lattice thermal conductivity.

\par With the finite-temperature phonons obtained, we proceed to calculate the lattice thermal conductivity of Cu$_3$BiS$_3$ using the Wigner transport equation~\cite{simoncelli2019unified, simoncelli2022wigner}, accounting for both phonon population and coherence contributions, as shown in Figure 3(c). Our results show that the thermal conductivity of the $P{2_12_12_1}$ phase at 400 K reaches 0.36 Wm$^{-1}$K$^{-1}$, in good agreement with the experimental value of 0.34 Wm$^{-1}$K$^{-1}$~\cite{maji2024three}, thereby validating our calculations. More specifically, at 400 K, the coherence contribution to the total thermal conductivity is 0.22 Wm$^{-1}$K$^{-1}$, accounting for \~61\% of the total thermal conductivity, highlighting the dominant role of phonon wave-like tunneling and strong lattice anharmonicity~\cite{yue2025interlayer}. We next continue to calculate the thermal conductivity of the $Pnma$ phase of Cu$_3$BiS$_3$ above 400 K using the Wigner transport equation~\cite{simoncelli2019unified, simoncelli2022wigner}(see in Figure 3(c)). Our theoretical model predicts an ultra-low total thermal conductivity of 0.36 Wm$^{-1}$K$^{-1}$, comprising 0.16 Wm$^{-1}$K$^{-1}$ from the population contribution and 0.20 Wm$^{-1}$K$^{-1}$ from the coherence contribution. Surprisingly, the predicted thermal conductivity of the $Pnma$ phase of Cu$_3$BiS$_3$ agrees well with the experimentally observed value of 0.34 Wm$^{-1}$K$^{-1}$ at 400 K for the partially occupied $Pnma$ phase. Furthermore, as temperature increases, the theoretical predictions remain in good agreement with experimental measurements for partial-occupied Cu$_3$BiS$_3$ ($Pnma$ space group), as shown in Figure 3(c). The cross-phase thermal transport in Cu$_3$BiS$_3$ remains unexplored~\cite{jia2023cu3bis3}, motivating us to compare the predicted thermal conductivities of both the $P{2_12_12_1}$ and $Pnma$ phases with experimental values, as shown in the inset of Figure 3(c). Our results show that the thermal conductivity of the $P{2_12_12_1}$ phase increases sharply above 400 K, deviating from the experimental values. In contrast, the $Pnma$ phase shows a continuous decrease in thermal conductivity above 400 K, showing a good agreement with experimental results at high temperatures. These results further confirm the temperature-induced phase transition from $P{2_12_12_1}$ to $Pnma$ and provide a clear explanation for the smooth change in thermal conductivity at the transition point. It is worth noting that the Wigner transport equation is limited by its reliance on the anharmonic heat flux contributions~\cite{simoncelli2022wigner}, and higher-order phonon scattering. Moreover, it cannot capture liquid-like ion diffuson~\cite{agne2018minimum}, such as the motion of ions in disordered or partially liquid environments. In contrast, molecular dynamics methods naturally include these effects, making them more suitable for such scenarios.

\begin{figure}[!htbp]
\centering
\includegraphics[width=13.5cm]{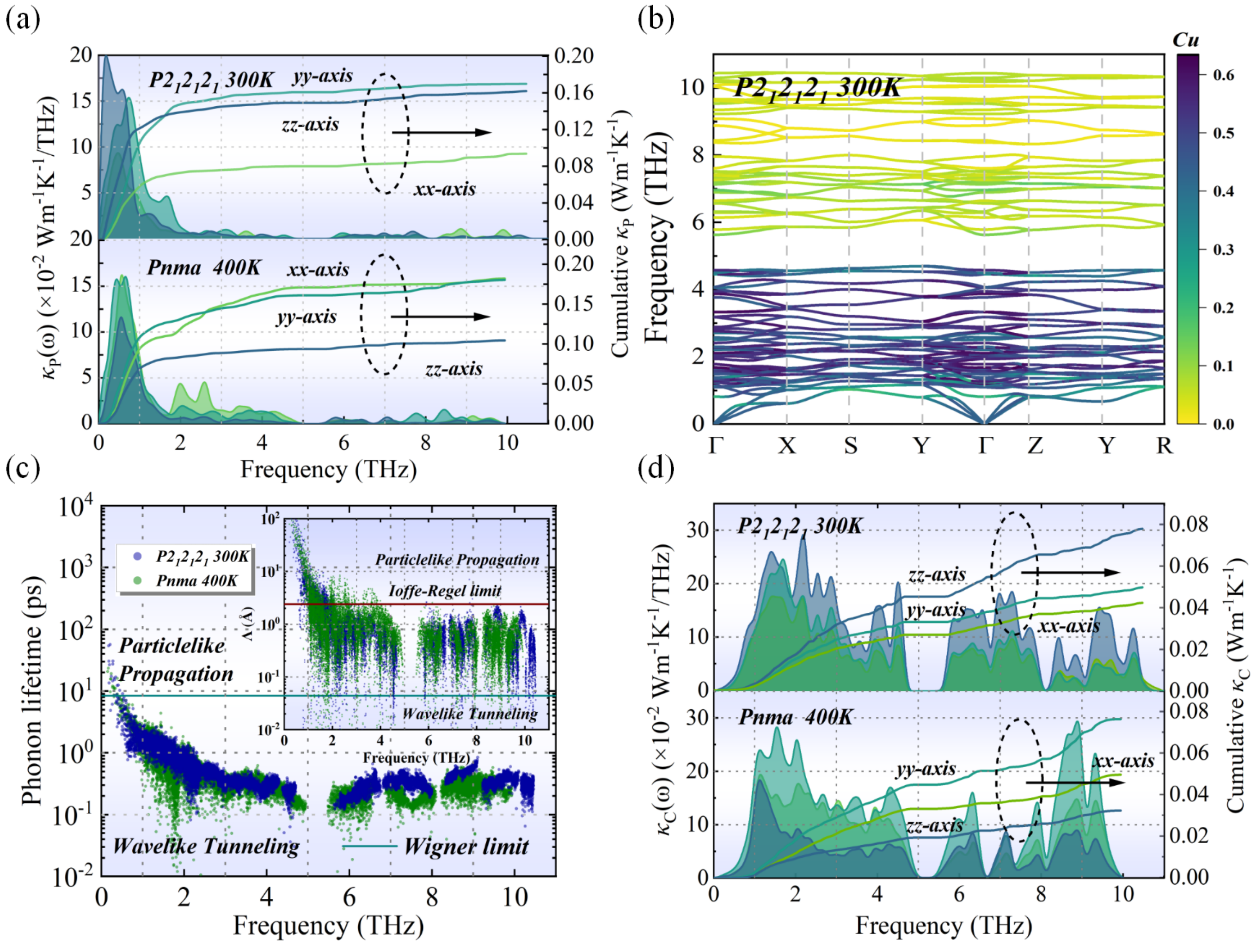}
\caption{(a) Calculated spectral/cumulative populations’ thermal conductivity $\kappa_P$ using the SCPB+3,4ph models along the three-direction for $P{2_12_12_1}$ phase at 300 K and $Pnma$ phase at 400 K, respectively. 
(b) Color-coded atomic participation ratio (APR) of Cu atoms projected onto the phonon dispersions for $P{2_12_12_1}$ phase at 300 K. 
(c) Calculated phonon lifetime as a function of frequency for $P{2_12_12_1}$ phase at 300 K and $Pnma$ phase at 400 K, where the green solid line represents the Wigner limit in time. Inset: Calculated phonon mean free path (MFP) as a function of frequency, where the red solid line represents the Ioffe-Regel limit in space. (d) Calculated spectral/cumulative coherences’ thermal conductivity $\kappa_C$ along the three-direction for $P{2_12_12_1}$ phase at 300 K and $Pnma$ phase at 400 K, respectively.}
\end{figure} 

\par Considering the liquid-like behavior of the $Pnma$ phase at high temperatures, we calculate $\kappa_L$ using the Green-Kubo equilibrium molecular dynamics (GK-EMD) method~\cite{kubo1957statistical} (see Figure S6 in the Supporting Information), which accounts for all terms in the heat flux operator~\cite{marcolongo2016microscopic, carbogno2017ab}. As shown in Figure 3(c), the total thermal conductivity predicted by the GK-EMD method at 400 K is in excellent agreement with that obtained from the Wigner transport equation~\cite{simoncelli2019unified, simoncelli2022wigner}. Even at 500 K, the discrepancy between the two methods remains negligible, with a difference of only 0.01 Wm$^{-1}$K$^{-1}$. Based on the above observation, and considering that the thermal conductivity from the Wigner transport equation~\cite{simoncelli2019unified, simoncelli2022wigner} is calculated for the $Pnma$ phase—i.e., in the absence of ion hopping—we conclude that ion hopping has minimal impact on thermal transport in liquid-like Cu$_3$BiS$_3$. The negligible effect of ion hopping on thermal transport can be attributed to the fact that ion-hopping behavior resembles rattling-like motion observed in certain crystalline materials~\cite{ghata2024exploring, tadano2018quartic}, which induces strong anharmonicity. This, in turn, suppresses phonon transport while enhancing the coherence contribution. Thus, in diffuson-dominated thermal transport regimes, ion hopping contributes minimally to the total thermal conductivity.

\par To further validate our theoretically predicted lattice thermal conductivity of crystalline Cu$_3$BiS$_3$, we compared the calculated values with experimental measurements in the temperature range of 2 to 100 K (see Figure 3(d)). Clearly, when boundary scattering at low temperatures is not considered~\cite{toberer2011phonon}, the predicted thermal conductivity—accounting only for three-phonon, four-phonon, and phonon-isotope scattering—deviates significantly from the experimental values. To accurately reproduce the experimental thermal conductivity~\cite{asheghi1997phonon, liu2004phonon}, effective grain sizes of 50, 100, and 200 nm are introduced, which significantly suppress the thermal conductivity and bring the predictions closer to the experimental results. On top of the 200 nm boundary scattering model, the inclusion of additional point-defect scattering leads to improved agreement between theory and experiment, as depicted in Figure 3(d). In addition, we performed a comparative analysis between the experimentally measured isobaric heat capacity (C$_p$) of Cu$_3$BiS$_3$ and the corresponding theoretical predictions within the temperature range of 2–70 K (see Figure S7 in the Supporting Information). Notably, at low temperatures, the experimental C$_p$ values exhibit excellent consistency with the calculated constant volume heat capacity (C$_v$), reflecting the reliability of the theoretical model in this regime. Meanwhile, our computational results closely reproduce the isobaric heat capacity data reported by Wu et al. prior to the observed phase transition~\cite{jia2023cu3bis3}.

\section{Microscopic mechanisms of thermal transport}

\par Next, to reveal the microscopic mechanisms of thermal transport in crystalline Cu$_3$BiS$_3$ with the $P{2_12_12_1}$ and $\mathrm{Pmna}$ phases, we calculate the spectral and cumulative propagating thermal conductivity at 300 K and 400 K, respectively. As shown in Figure 4(a), phonons with frequencies below 1.5 THz dominate the particle-like phonon channel, which can be attributed to the relatively large group velocities of the acoustic and low-frequency optical modes (see Figure S8 in the Supporting Information). To elucidate the correlation between the ultra-low propagating thermal conductivity and the atomic structure of crystalline Cu$_3$BiS$_3$, we project the atomic contributions onto the phonon bands, as shown in Figure 4(b). Specifically, Cu atoms predominantly contribute to the low- and mid-frequency optical phonon modes in the range of 1.5–5.0 THz. In contrast, S atoms primarily participate in the high-frequency optical modes above 5.0 THz, while Bi atoms exhibit relatively low participation and are mainly involved in a limited number of low-frequency optical modes between 1.5 and 2.0 THz (see Figure S9 in the Supporting Information). The fully Cu- and Bi-occupied low-frequency optical modes lead to strong phonon scattering rates, thereby suppressing thermal transport in crystalline Cu$_3$BiS$_3$, a phenomenon similarly observed in CsCu$_4$Se$_3$~\cite{yue2025interlayer}. Importantly, Cu atoms give rise to low-energy, Einstein-like quasilocalized vibrational modes~\cite{tadano2018quartic}, particularly within the 2–4 THz frequency range. These Cu-dominated quasilocalized vibrational phonon modes act as strong phonon scatterers, significantly interacting with and disrupting the propagation of heat-carrying acoustic phonons. This phenomenon has also been reported in cubic SrTiO$_3$\cite{tadano2015self}, the double perovskite Cs$_2$AgBiBr$_6$\cite{zheng2024unravelling}, and copper-based chalcogenides such as \textit{o}-CsCu$_5$S$_3$\cite{yue2024ultralow}.

\par As illustrated in Figure 4(c), we further calculate the phonon lifetimes for the $P{2_12_12_1}$ phase at 300 K and the $Pnma$ phase at 400 K, respectively. We find that the majority of phonon modes exhibit lifetimes shorter than the Wigner limit~\cite{simoncelli2019unified, simoncelli2022wigner}, considering both three-phonon (3ph) and four-phonon (4ph) scattering processes, indicating strong anharmonicity in the system. For the $P{2_12_12_1}$ phase, we observe a dip in phonon lifetimes for phonons around 0.75 THz, which are dominated by Bi atoms, highlighting a strong scattering source that contributes to the suppression of thermal conductivity~\cite{zheng2022anharmonicity, zheng2024unravelling}. Similarly, the flattened, localized vibrational mode, primarily associated with Cu atoms near 2.2 THz, also exhibits an extremely short lifetime and pronounced anharmonic scattering (see Figure 4(c) and Figure S10 in the Supporting Information). Therefore, we further confirm that the ultra-low thermal conductivity is attributed to the vibrations of Cu and Bi atoms. In the $Pnma$ phase, Cu atoms play a similar role, as the Cu-dominated mode near 1.8 THz exhibits a significant drop in phonon lifetime (see Figure 4(c) and Figure S10 in the Supporting Information), corresponding to the dip observed in the spectral thermal conductivity (see Figure 4(a)). Although strong anharmonicity and the low phonon lifetimes of quasilocalized modes suppress particle-like phonon propagation~\cite{mukhopadhyay2018two}, they enhance the wave-like phonon diffuson contribution (see Figure 4(d)). Theoretically, the presence of dense, quasilocalized phonon modes with similar energies and strong anharmonicity can facilitate significant heat exchange between phonon modes in energy space~\cite{dangic2025lattice, yue2025interlayer}, thereby promoting diffuson-like transport~\cite{caldarelli2022many}.
 
\begin{figure}[t]
\centering
\includegraphics[width=14cm]{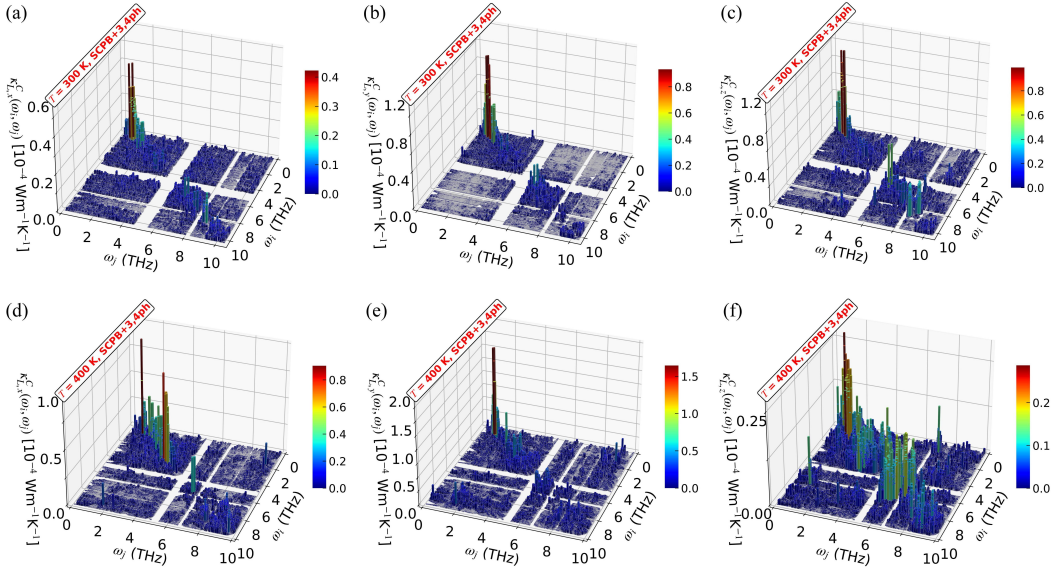}
\caption{(a) Three-dimensional visualizations of the coherences' thermal conductivity $\kappa_C$($\omega_{qj}$,$\omega_{qj'}$) based on the SCPB+3,4ph model along with the $x$-axis for the $P{2_12_12_1}$ phase at 300 K. The diagonal data points ($\omega_{qj}$ = $\omega_{qj'}$) indicate phonon degenerate eigenstates. (b) The same as (a), but for $y$-axis. (c) The same as (a), but for $z$-axis. (d) The same as (a), but for the $Pnma$ phase at 400 K. (e) The same as (d), but for $y$-axis. (f) The same as (e), but for $y$-axis.}
\end{figure}

\par To gain a better understanding of microscopic mechanisms behind the coherences’ thermal conductivity $\kappa_C$ in Cu$_3$BiS$_3$, we calculate the spectral and cumulative mode-specific contributions to coherences’ conductivity at 300 and 400 K, respectively, as shown in Figure 4(d). In sharp contrast to the spectral populations’ conductivity $\kappa_C$ of Cu$_3$BiS$_3$ in Figure 4(b), the majority of the coherences’ conductivity from the wavelike tunneling channel is carried by the phonons with almost full-frequency at both 300 and 400 K. This can be attributed to the small inter-band spacing within full-frequency region (due to dense phonon dispersions (see Figure S11 in the Supporting Information)) in conjunction with large linewidths (large scattering rates (see Figure 4(c)) or strong anharmonicity. Specifically, the spectral coherence conductivity exhibits a pronounced peak at 2.2 THz in the $P{2_12_12_1}$ phase and at 1.8 THz in the $Pnma$ phase. These peaks correspond to Einstein-like quasilocalized vibrational modes predominantly associated with Cu atoms (see Figure 4(b)). Within the middle and high-frequency domains, the contribution from spectral coherences' conductivity is considerable, largely driven by strong anharmonic scattering alongside small interband spacing. Overall, for the Cu$_3$BiS$_3$,strong anharmonic effects combined with reduced phonon group velocities lead to quasilocalization in real space, thereby suppressing phonon-gas transport and fostering strong phonon coupling alongside significant diffuson contributions. Figures 5(a-f) show that the contributions to coherences’ thermal conductivity in Cu$_3$BiS$_3$ at 300 and 400 K, can be resolved in terms of the phonon energies, namely, $\omega_{qj}$ and $\omega_{qj'}$ of two coupled phonons. At 300 K, similarly to the harmonic glasses~\cite{simoncelli2023thermal}, where quasi-degenerate phonon states dominate (see Figures 5(a-c)), the phonons contributing to $\kappa_C$ along the three axe have relatively the same frequencies. For the $Pnma$ phase at 400 K, the two-dimensional density of states for the coherences’ thermal conductivity along the $x$-aixs and $y$-aixs (see Figures 5(d-e)) also shows couplings between quasi-degenerate states. In contrast, the the coherences’ thermal conductivity along the $z$-aixs involves couplings between phonon modes with significantly different frequencies (see Figure 5(f)), driven by strong anharmonicity. As a result, the associated heat transport mechanism is fundamentally different from that in harmonic glasses.

\section{Conclusion}
\par In summary, we have comprehensively investigated the anharmonic lattice dynamics and thermal transport behavior in both ordered and partially disordered Cu$_3$BiS$_3$ using a combination of experimental measurements and advanced DFT-based theoretical simulations. To begin with, we experimentally synthesized and characterized both the ordered structure stable at room temperature and liquid-like structure occurring above 400 K of Cu$_3$BiS$_3$. The latter disordered structure crystallizing in the $Pnma$ space group, is characterized by statistical distribution of Cu+ ions on several crystallographic sites and emerges upon temperature increase - an observation further supported by ab initio molecular dynamics simulations. Subsequently, we employ DFT-based self-consistent phonon theory with bubble-diagram corrections, combined with the Wigner transport equation, to capture the anharmonic lattice dynamics and uncover the microscopic mechanisms underlying the ultra-low thermal conductivity in Cu$_3$BiS$_3$. 
At low temperatures (2-100 K), the predicted thermal conductivity-accounting for 3ph, 4ph, isotope, 200 nm boundary, and point-defect scattering, accurately reproduce the experimental thermal conductivity.
At 400 K, our results demonstrate that both the ordered $P{2_12_12_1}$ phase and disordered $Pnma$ phase exhibit ultralow thermal conductivity, with values of 0.36 Wm$^{-1}$K$^{-1}$. Notably, our predicted thermal conductivity, incorporating both propagating and diffuson-like phonon contributions, well reproduces the experimental measurements of 0.34 Wm$^{-1}$K$^{-1}$, even for the partially Cu-occupied $Pnma$ phase. In both the $P{2_12_12_1}$ and $Pnma$ phases, the ultralow thermal conductivity is mainly due to strong anharmonic scattering associated with Bi and Cu atoms. In particular, large anisotropic vibrations of Cu in triangular coordination leads to low-energy, Einstein-like vibrational modes that efficiently scatter heat-carrying acoustic phonons, thereby hindering their particle-like propagation. On the other hand, machine-learning-based Green–Kubo molecular dynamics simulations further confirm that ion hopping has a negligible impact on heat transport, even in the high-temperature, liquid-like phase. Therefore, despite the disordered nature of Cu$_3$BiS$_3$, our models—both machine learning and first-principles-based—provide reasonable estimates of thermal conductivity when using ordered structural representations. Our findings on Cu$_3$BiS$_3$ suggest that the lattice thermal conductivity ($\kappa_L$) of amorphous or disordered compounds can, to a reasonable degree of accuracy, be approximated using calculations based on their ordered crystalline counterparts. This approach provides a practical pathway for evaluating thermal transport properties in complex systems where fully capturing structural disorder remains computationally challenging. Moreover, our study highlights the minimal impact of ionic mobility on thermal conductivity and offers valuable insights into the fundamental phonon transport mechanisms underlying the ultralow $\kappa_L$ observed in both the ordered and liquid-like phases of Cu$_3$BiS$_3$.

\section*{Acknowledgements}
J.Y. acknowledges the National Natural Science Foundation of China through grants No. 52072188 and No. 12204254, the Program for Science and Technology Innovation Team in Zhejiang through grant No. 2021R01004, as well as the Institute of High-pressure Physics of Ningbo University for its computational resources. X.S. acknowledges funding supported by National Youth Talent Project and the Fundamental Research Funds for the Central Universities (D5000250021), and the European Union's Horizon 2020 research and innovation program under the Marie Sklodowska-Curie grant agreement No 101034329 and the WINNING Normandy Program supported by the Normandy Region. X.S, K.M, and E.G gratefully thank Jean-François Lohier, Fabien Veillon, Christelle Bilot, and Jerôme Lecourt for technical support. S.L. acknowledges the computation resources provided by the National Academic Infrastructure for Supercomputing in Sweden (NAISS) -- at supercomputer centers NSC and PDC, partially funded by the Swedish Research Council through grant agreement no. 2022-06725 -- and by the Vienna Scientific Cluster (VSC) in Austria.

\section*{Conflict of interest}
The authors declare that they have no known competing financial
interests or personal relationships that could have appeared to influence the work reported in this paper.


\printendnotes

\bibliography{ref}

\begin{biography}[example-image-1x1]{A.~One}
Please check with the journal's author guidelines whether author biographies are required. They are usually only included for review-type articles, and typically require photos and brief biographies (up to 75 words) for each author.
\bigskip
\bigskip
\end{biography}

\clearpage
\graphicalabstract{example-image-1x1}{Please check the journal's author guildines for whether a graphical abstract, key points, new findings, or other items are required for display in the Table of Contents.}

\end{document}